**Classification:**
    **Major:** Physical sciences
    **Minor**: Chemistry, Physics, Applied Physical Sciences, Applied Mathematics

# A Precise Packing Sequence for Self-Assembled Convex Structures

Ting Chen[1*], Zhenli Zhang[1*] and Sharon C. Glotzer[1,2,†]

[1]Department of Chemical Engineering and [2]Department of Materials Science & Engineering

University of Michigan, Ann Arbor, Michigan 48109-2136

---

[*] These authors contributed equally to this work.
[†] To whom correspondence should be addressed. Email: sglotzer@umich.edu

**Abstract**

Molecular simulations of the self-assembly of cone-shaped particles with specific, attractive interactions are performed. Upon cooling from random initial conditions, we find that the cones self assemble into clusters and that clusters comprised of particular numbers of cones (e.g. 4 - 17, 20, 27, 32, 42) have a unique and precisely packed structure that is robust over a range of cone angles. These precise clusters form a sequence of structures at specific cluster sizes– a "precise packing sequence" -- that for small sizes is identical to that observed in evaporation-driven assembly of colloidal spheres. We further show that this sequence is reproduced and extended in simulations of two simple models of spheres self-assembling from random initial conditions subject to certain convexity constraints. This sequence contains six of the most common virus capsid structures obtained *in vivo* including large chiral clusters, and a cluster that may correspond to several non-icosahedral, spherical virus capsid structures obtained *in vivo*. Our findings suggest this precise packing sequence results from free energy minimization subject to convexity constraints and is applicable to a broad range of assembly processes.

The recent fabrication via evaporation-driven self-assembly of colloidal clusters containing micron-sized plastic spheres arranged in perfect polyhedra (1) has generated much excitement in the materials community. Precise structures such as these are rare in materials self-assembly problems (2, 3), with some notable, recent exceptions (e.g., (4) (5) (6)). In biology, however, precise structures are commonplace. A classic example is that of viruses, in which protein subunits, or small groups of protein subunits called capsomers, self-organize into precisely structured shells with icosahedral and other symmetries (7). Although obtained via wholly different processes, the precise packing of colloidal clusters (often referred to as colloidal "molecules" (8) and virus capsids motivates us to investigate, in this paper, the minimum set of organizing principles required for the self-assembly of precise structures such as these.

One interesting building block shape that self-organizes into precise structures is the cone. Certain amphiphilic nanoparticles (9), molecules (4, 10, 11) and some virus capsomers (12, 13) that self-assemble into precise structures can, to first approximation, be modeled as cone-shaped particles associating via weak attractive interactions. What are the rules that govern the self-assembly of finite numbers of cones into precise clusters? Tsonchev *et al*. (14) presented a geometric packing analysis to explain the spherical shape resulting from the self-assembly of $N$ cone-shaped amphiphilic nanoparticles in the limit of very large $N$, in which local hexagonal packing of hard cones is assumed. For clusters of smaller numbers of particles, however, the finite curvature of the assembled cluster will prevent extended hexagonal order (15), altering the cluster shape in an unknown way.

Here we investigate the clusters formed via self-assembly of attractive cone-shaped particles using Monte Carlo simulation. We find that, by varying the cone angle, the cones form a sequence of precise, convex structures that are reproduced and extended by simulations of simple models of spheres self-assembling under certain convexity constraints. Remarkably, we find that this packing sequence successfully describes *both* the polyhedral structures formed by colloidal particles self-assembling on an evaporating droplet, and icosahedral virus capsid structures formed from protein capsomers, thereby elucidating fundamental organizing principles common to the structures in these disparate systems.

The paper is organized as follows. First we describe how we model the self-assembly of cones with specific, attractive interactions and present the details of the structures we obtain from simulation. We then hypothesize a minimum set of organizing principles to explain the notable resemblance between the polyhedral structures formed by cones and the clusters formed by colloidal spheres in an evaporation process. Two minimal models of spheres are proposed and simulated to verify this hypothesis. We next compare the simulated structures possessing icosahedral symmetry with the known icosahedral virus capsid structures, and discuss the implications of our simulation results in the context of current theoretical and experimental progress on virus capsid assembly. Detailed descriptions of the methods we use are presented in the Methods section.

## Results and Discussion

**Self-assembly of cone-shaped particles.** We carry out Monte Carlo (MC) simulations (see Methods section) on collections of cone-shaped particles, each comprised of a linear array of overlapping hard spherical beads of decreasing size connected rigidly together (Fig. 1). Interactions between cones are specific, attractive, and occur bead-wise, and are designed to promote alignment of cones, thereby facilitating self-assembly of clusters that would be difficult to obtain with non-specific cone-cone interactions using MC simulation. Our model "patchy (16)" cones resemble nanocolloidal particles that have been synthesized in the shape of hollow cones (17) and ice cream cones (18, 19), and may soon be fabricated using new methods for producing multiphasic particles (20).

We systematically vary the cone angle and study the size and structure of the self-assembled clusters formed on cooling from an initially random configuration. We find that for a given cone angle, a distribution of cluster sizes is observed; for a few angles, the distribution is narrow and highly monodisperse, yielding essentially only one cluster size. Such is the case for clusters containing 4, 6, and 12 cones, corresponding to cone angle ranges 102.5° - 116.4°, 81.1° - 92.1°, and 57.4° - 62.7°, respectively. For other angles a relatively broad distribution of cluster sizes is observed. Interestingly, we find that the structure of certain cluster sizes is robust and independent of angle, although the yield may vary with angle. Specifically, we find by visual inspection that clusters sizes of $N = 4 – 17, 20, 27, 32$ and 42 have a unique and precise structure that is independent of cone angle (we note that, at $N = 8$ two isomeric clusters are found, a snub-disphenoid at the largest cone angle and a twisted-square structure at the smallest cone angle). All other cluster sizes between 17 and 42 not listed here exhibit structures that are not robust, and can have different structures and/or symmetries for different clusters of the same size $N$. Thus, e.g. a cluster of 20 particles obtained for a cone angle of 45.3° is structurally identical to a cluster of 20 particles obtained for slightly larger or smaller angle, whereas the same is not true for a cluster of 21 particles. Details regarding the dependence of cluster size and structure on cone angle will be reported elsewhere (21). Here, we focus on the "packing sequence" formed by those robust, precise clusters whose structure is independent of cone angle. "Magic number" clusters are well known for dilute systems of particles interacting via a Lennard-Jones interaction (22), but those clusters differ from the clusters obtained here in that the clusters of cones are all convex (that is, the outermost bead of each cone sits on a convex hull).

Fig. 2a shows the simulated "magic number" clusters ranging from $N = 4$ to 17. Remarkably, we find that the simulated structures are identical to those reported in experiments on the evaporation-driven self-assembly of polystyrene (PS) colloidal spheres (1) for $N = 4 - 15$ with one exception at $N = 11$, where a concave structure was observed experimentally. A subsequent study by the same group on polymethylmethacrylate (PMMA) colloidal particles (23) instead found a convex structure at $N = 11$, the same as that predicted for our cone-shaped particles. This suggests that the concave structure observed for $N = 11$ in the original experiment forms because the PS particles provide insufficient support to maintain the final structure's convex shape. Additionally, the snub-disphenoid structure of the 8-cone cluster was observed experimentally in references (1) and (23) whereas the twisted-square structure was observed in the experiments reported in references (24, 25).

**Self-assembly of spherical particles (small $N$).** The observation that our cone-shaped particles self-assemble into the same convex structures as those observed in the evaporation experiment is nontrivial since the two self-assembly processes are driven by different mechanisms: the model cones by highly specific, anisotropic attractions and the colloidal spheres by capillary forces. That both processes result in the same packing sequence suggests a more general underlying physical principle common to both. What might this common principle be? First, both self-assembly processes occur via the minimization of free energy. Second, a salient feature of this packing sequence is the convex shape of the structures. In the system of cone-shape particles, the convex shape of the shell formed by the outermost beads arises from the directional attraction between cones and conical excluded volume which pushes the head of each cone to a convex surface, while in the evaporation-driven assembly of colloidal spheres the convex shape likely results from both the tendency of the particles to sit at the droplet interface as the droplet shrinks, and the contact force between the particles. Therefore, we postulate that the unique packing sequence may arise from free energy minimization subject to a convexity constraint.

To test our conjecture we study two simple models that each contains a finite number of hard spheres of diameter $\sigma$ which self-assemble subject to a convexity constraint. Two forms of potential energy are investigated to ascertain the dependence, if any, of cluster structure on the choice of driving force for self-assembly. In one system the spheres interact with a short-range, square-well attraction ($\lambda = 0.2\sigma$) in addition to the hard-core repulsion and convexity constraint. In the second system the potential energy has the form of the second moment of the mass distribution $U = \sum_{i=1}^{N} |\mathbf{r}_i - \mathbf{r}_0|^2$ where $\mathbf{r}_i$ is the position vector of particle i and $\mathbf{r}_0$ is the centroid of the cluster, again in addition to a hard-core repulsion and convexity constraint. In the second system, particles are not attracted directly to each other but are instead attracted to the cluster center of mass at all times. Our motivation for choosing these two potentials is that (i) the cones interact with a square well potential similar to the first sphere model, and (ii) the colloids in the evaporation experiments move under a linear capillary force directed, presumably, towards the center of the evaporating droplet (1, 26) resulting in a quadratic potential of the type proposed for the second sphere model. As suggested in Ref. (26) the potential energy of the colloidal system can be estimated from a sum over all particles of the product of the net force on the particle and the distance of the particle to the droplet center, which is a quadratic function and has the form of the second moment of the mass distribution, $M = \sum_{i=1}^{N} |\mathbf{r}_i - \mathbf{r}_0|^2$, with a prefactor. Mathematical solutions for "minimal second moment of mass distribution" structures have been reported in the literature (27), but not subject to a convexity constraint.

We perform MC simulations with a convex-hull searching algorithm to study the assembly of the two model sphere systems. The convex-hull algorithm imposes a requirement on the convexity of the assembled structures that constrains all particles to the surface of the convex hull formed by the particles at all times. In this way, concave clusters, such as those with particles in the center, are forbidden. Details are reported in the Methods section.

The simulated structures for $N$ = 4 to 17 are shown in Fig. 2b. The packing sequences obtained for the two sphere models are identical, and also identical to those formed by the cones (shown in Fig. 2a) and colloids assembled by evaporation (23).[§] This result supports our conjecture that the packing sequence results simply from free energy minimization subject to a convexity constraint since systems with different potential energy functions produce identical packings. We calculate the values of the second moment of the mass distribution for our simulated structures and compare them with previous reported experimental and simulation results. The average relative difference of the second moments of mass distribution between our simulated structures and the experimental structures is 0.2%, in contrast to an average relative difference of 0.9% between experimental results and previously reported numerical simulation results, for $N$ = 4 - 14 (26). Nearly the same packing sequence has been seen in other experiments on colloidal clusters using different types of particles (24, 25), which further supports the robustness of this packing sequence. To our knowledge the structures we obtain at $N$ = 16 and 17 have not been reported by any experiment or theory, and thus are presented here as a prediction.

---

[§] Note that in the minimal model simulations, only the snub-disphenoid structure is observed at $N$ = 8.

**Self-assembly of cones and spherical particles (moderate to large $N$).** Fig. 3ab shows the self-assembled clusters of cones we obtain for $N$ = 20, 27, 32 and 42 and their identical spherical particle counterparts. As noted above, clusters of size $N$ between these four values either lack precision or lack high symmetries and are not robust or independent of cone angle, in contrast to these four cluster sizes. The shape of these larger clusters is consistent with the theoretical prediction of Tsonchev *et al.* that spherical structures are preferred in the limit of large $N$ (14). To expedite the simulation and facilitate assembly for $N > 42$, and motivated by the expected roughly spherical shape of large clusters, we reduce the available phase space by imposing "spherical convexity" *in the initial stage only* of the simulations of the sphere models. To do this, we randomly distribute the spherical particles on the surface of a large fictitious sphere, which is then gradually reduced in diameter to bring the particles together, as done by Zandi *et al.* (28) in their study of the self-assembly of 2-D disks in the context of virus capsid formation. In our simulations, however, eventually the "guiding" sphere is withdrawn and the particles continue self-assembling under solely the constraint of convexity, rather than the more stringent constraint of spherical convexity. In this way, for both the "square-well" and "second moment" model sphere systems, we successfully reproduce the packing sequence in both Fig. 2 and 3b, and we are able to predict precise packings, including those with both right- and left-handed chiral structures (see below), for substantially larger $N$, as shown in Fig. 3b for $N = 50$, 72, and 132.

What insights can be gleaned from the importance of spherical convexity on guiding the initial stages of assembly for these moderately sized clusters? We find that there exists a critical number density $0.87 < n_c < 0.89$, defined by $n_c = \dfrac{N}{4\pi R^2}$ where $R$ is the distance of the spherical surface's center to the center of spherical particles on the surface, at which the spherical convexity constraint can be relaxed. This critical density is in the neighborhood of the estimated freezing density[§], the density at which the assembly first exhibits solid-like behavior, suggesting that the particles are close enough to begin to order on the spherical surface just before the spherical convexity constraint is removed. Further comparing the configurations immediately before and after the critical density is reached, we find that the symmetry of the final packing appears just prior to $n_c$ for all structures except $N = 132$. This finding indicates that for large $N$, self-assembly may occur in two key stages. In the first stage, particles arrange into a specific symmetry and in the second stage, particles arrange with that symmetry into a precise polyhedral shape. For small $N$, only one stage of assembly appears to be required, as no initial "guiding sphere" is needed to achieve the observed convex structures. This hypothesis is further supported by our recent findings, presented elsewhere, that imposing an initial constraint of prolate

---

[§] The estimated values of the freezing density for $N = 32$, 42, 50 and 72 are $n_c = 0.891$, 0.904, 0.917, and 0.929 for square-well hard sphere clusters and $n_c = 0.815$, 0.840, 0.891, and 0.891 for minimal second moment clusters. We can compare these critical densities to the freezing densities of hard disks on a spherical surface ($n_c = 0.87$) (15) and a flat surface ($n_c = 0.89$) (29, 30) and square-well hard disks and hard spheres with $\lambda = 0.20\sigma$ ($n_c = 0.808$) in 2-D (31) all of which are fairly close. We thus take $0.87 \sim 0.89$ as a rough estimate of the freezing density for our system of square-well hard spheres, keeping in mind that this quantity is calculated in the limit of a large spherical surface, or a flat surface.

spheroidal convexity produces polyhedral clusters resembling prolate virus capsid structures (32).

**Comparison with virus capsids.** It is instructive to compare our predicted packing sequence to self-assembled viral capsids since both share the same icosahedral symmetry and shape for certain $N$. The origin of the precise packing of protein capsomers in a virus capsid shell has been the focus of much recent theoretical work (28, 33-39). About half of all viruses found in humans, animals and plants are sphere-like, and most of them have protein shells possessing icosahedral symmetry. With a few notable exceptions, virus capsids contain capsomers that consist of a few distinct protein subunits, and exist only in special $T$-number (triangulation number, $T = 1, 3, 4, 7, 13$, etc., which refers to the number of distinct environments for protein subunits) structures of Caspar-Klug (CK) "quasi-equivalence" theory (40), so named for the non-identical but similar ("quasi-equivalent") environments of protein subunits in virus capsids described by the theory. The CK theory assumes that all icosahedral viruses have $60T$ subunits arranged into 12 pentamers (capsomers consisting of five protein subunits) and $10(T-1)$ hexamers (capsomers consisting of six protein subunits) and views virus assembly as a free-energy minimizing, crystallization process. Very recently, CK theory was shown to be a subset of a more general theoretical framework based on tiling theory (38). Some large virus capsids are chiral. For example, murine polyomavirus (41) and simian virus 40 (SV40) (42) both have $7d$ (*dextro-*, or right-handed) form, while bacteriophage HK 97 (43) has $7l$ (*laevo-*, or left-handed) form. The outer shell of the double-shelled rice dwarf virus (44) and rhesus rotavirus (45) has $13l$ symmetry and Bursal disease virus (46) has $13d$ symmetry.

The structures in the packing sequence presented here at $N = 12, 32, 72$ and $132$ are identical to the icosahedral structures found in $T = 1, 3, 7$, and $13$ virus capsids, respectively. Even right-handed and left-handed chiral configurations at $N = 72$ ($T = 7$) and $N = 132$ ($T = 13$) are generated successfully; to our knowledge, this is the first simulation of the $T = 7l, 13l$ and $13d$ virus capsid structures. Moreover, our packing sequence accounts for the existence of non-quasi-equivalent virus capsid structures. For example, the Polyoma virus (47), which contains 72 identical pentamers and consists of 360 total subunits, should form a $T = 6$ structure according to the CK quasi-equivalence theory. Instead, it forms a $T = 7d$ structure *in vivo*, as predicted by the tiling theory of Twarock and coworkers (38). Our model successfully predicts the $T = 7d$ icosahedral structure for $N = 72$. Additionally, the cluster structure found for $N = 27$ may correspond to several non-icosahedral spherical virus structures, such as the middle component of the pea enation mosaic virus, the top component of the tobacco streak virus and the Tulare apple mosaic virus, which consist of about 150 subunits or 27 capsomers (12 pentamers and 15 hexamers) and violate CK quasi-equivalence theory, as suggested by Cusack (48). To our knowledge, this is the first theoretical or computational prediction of this experimentally-proposed structure.

Notably, the cluster structure we obtain for $N = 42$ does not have the expected icosahedral symmetry predicted by CK theory and observed experimentally for the $T = 4$ virus shell structure. Indeed, we did not observe an icosahedron for $N = 42$ in any of over a hundred simulation runs. Moreover, simulations beginning with an $N = 42$ cluster artificially arranged into the expected icosahedral symmetry evolved to the observed non-icosahedral structure. Using two types of spherical particles with a size ratio of 1 : 0.8 and longer range attraction (28) via a Lennard-Jones potential, we are able to obtain the icosahedral structure at $N = 42$ with small probability. However, the prevalence of the non-icosahedral structure in our simulations under all conditions and the evolution of the structure away from icosahedral symmetry indicates that icosahedral symmetry is not a free energy minimum at $N = 42$. Interestingly, the symmetry of this non-icosahedral $N = 42$ structure is also predicted in the proposed optimal packings of the maximum volume of a convex hull for a set of points on a sphere (49) – a closely related mathematical problem – and in the packings of point charges on a sphere, known as the Thomson problem in mathematics (50). Apparently, additional types of inter-particle interactions or additional constraints are required to obtain an icosahedral structure for $N = 42$. However, the ability of the simple models presented here to obtain other complex capsid shapes argues for the general utility of this phenomenological approach for illuminating certain fundamental organizing principles at work in the problem of virus assembly.

**Relation to previous simulation studies of virus assembly.** Further discussion on the potential relevance to virus capsid assembly is warranted. Several theoretical models have explored the underlying mechanism of virus capsid assembly (28, 33-38). Most models strictly impose spherical geometry and/or icosahedral symmetry at all times, and the simplest models generate unwanted multiplicity or degeneracy of structures not observed in nature. Our minimal model of spheres self-assembling under first a spherical, and then a general, convexity constraint may be viewed as a phenomenological model for icosahedral virus assembly. In our model, no *a priori* assumption on the shape or symmetry of the packing is made for small cluster sizes, nor for larger cluster sizes after the critical density is reached, yet the unwanted multiplicity of structures does not occur for the sequence of structures presented. Recently, Zandi *et al.* showed that a unique, minimum energy icosahedral structure formed from a collection of identical disks self-assembling on a spherical surface is greatly facilitated by either a small compression of the capsid, which they achieved by slightly shrinking the spherical surface to induce overlap between the assembling units, or by using two different size units (28). Both of these mechanisms may be viewed as different ways of relaxing the constraint of spherical convexity while still enforcing general convexity, and thus we can view these approaches as special cases of the more general approach investigated here.

How might this initial spherical convexity constraint arise in virus assembly? It has been argued (51) that the interaction between protein units and nucleic acids induce the proteins to assemble into a spherical structure around the genomic material as a first step in capsid formation. As the nucleic acid is neutralized by the amino terminal tails on the assembling proteins, the genomic material shrinks, pulling the capsid proteins in tighter until they eventually restructure into their ultimate icosahedral symmetry (51). The existence of a critical packing density in our simulations that appears to delineate two stages in the assembly process (34, 51) may support this conjecture. In general, a convexity constraint --- which we find to be the fundamental constraint needed to obtain precise structures reliably for the range of cluster sizes reported --- may arise from internal pressure created from the genetic material enclosed by the capsid, (28) or from the scaffold proteins, or from excluded volume packing constraints and hydrophobic/hydrophilic interactions among the capsomers.

## Concluding Remarks

We have considered the clusters formed via self-assembly of attractive cone-shaped particles using Monte Carlo simulation. We showed that, by varying the cone angle, the cones form a sequence of precise, convex structures that are reproduced by simulations of minimal models of spheres self-assembling under a general convexity constraint for $N = 4 - 17$. The sequence is shown to be identical to that obtained in experiments on evaporation-driven colloidal assembly for small cluster sizes. For moderate to large cluster sizes, the addition of an initial spherical convexity constraint in the minimal model simulations of spheres extends the sequence to produce several known icosahedral virus capsid structures. Inclusion of the spherical convexity constraint in the initial stages of formation of small $N$ clusters produces identical packings as without the spherical constraint. Our findings demonstrate that interactions that induce particle clustering combined with certain convexity constraints lead to a seemingly universal sequence of precise, robust packings. Barring kinetic traps, these features appear to be necessary and sufficient conditions for achieving the packing sequence presented here. This result provides an unexpected link between seemingly disparate phenomena –- evaporation-driven assembly of polyhedral clusters from micron-sized colloidal particles, self-assembly of polyhedral clusters from cone-shaped particles with directional attractions, and self-assembly of nanometer-sized virus capsids from protein capsomers –- and shows them to be members of the same packing sequence.

**Methods**

The cone model we use is shown in Fig. 1. Within a single cone, the distance between neighboring beads is 0.5 in units of σ, where σ is the diameter of the smallest bead. The cone angle $\theta$ is defined by the relation $\sin(\frac{\theta}{2}) = \frac{r_6 - r_1}{2.5}$, where $r_6$ and $r_1$ are the radii of the largest and smallest beads in the particle, respectively. The constant 2.5 is the distance between the centers of the two end beads. Square-well interactions of well depth $\varepsilon$ exist between beads occupying the same positions (indicated by color) on the particles except for the two end beads, and for unlike beads, which interact with other beads through hard-core excluded volume only. The interaction range (width of square well) $\lambda = 0.4\sigma$.

To study the self-assembly of cone-shaped particles we use Monte Carlo simulation in the canonical (constant number of particles, constant volume, constant temperature) ensemble, in which systems containing hundreds of particles are cooled from an initially disordered configuration in three dimensions to form stable structures. Multiple independent runs are performed and slow cooling and/or annealing is used to avoid kinetically trapped structures. All systems start from a high temperature, disordered state containing $M = 200$-$1000$ cone-shaped particles of angle $\theta$. Each system is gradually cooled from high temperature ($k_B T/\varepsilon = 100.0$) to a target temperature of $T = 0.3$-$0.5$ (all parameters are in reduced units). At each iteration, a randomly chosen particle attempts to rotate or translate by a small amount subject to the standard Metropolis importance-sampling algorithm. The MC moves consist of moves on the spherical surface (52) and either radial moves for the square-well hard sphere system or random translational moves for the "second moment" system. The ratio of radially inward to radially outward moves = 9 : 1, and the ratio of radial or translational moves to surface moves = 2 : 8. Many tens of millions of iterations are needed to fully assemble the largest clusters. Multiple independent runs are performed with different initial configurations and along different cooling paths to avoid local minimum free energy structures and achieve equilibrium structures.

To simulate the self-assembly of the attractive hard sphere models subject to convexity constraints, we implement the "incremental algorithm" (53) from computational geometry to identify the convex hull of our structures as they assemble on-the-fly in our simulations. The spheres are initially randomly distributed on a large convex surface and subsequently moved subject to the convexity constraint defined above. Any trial moves resulting in a structure violating the convexity criterion are rejected and trial moves are either rejected or accepted according to the conventional Metropolis importance-sampling algorithm. For simulations corresponding to the precisely structured clusters, this approach reliably gives the same cluster structure for each energy minimization.

We thank Prof. I. Ilinkin and Prof. E. Gottlieb at Rhodes College for helpful information on the convex hull algorithm, and Prof. G. Huber of the University of Connecticut and Prof. R. Ziff of the University of Michigan for helpful comments on the critical packing density. We also thank Mr. X. Zhang in the Glotzer group at the University of Michigan for programming assistance and Prof. R.G. Larson of the University of Michigan for a careful reading of the manuscript. Financial support was provided by the National Science Foundation, under Grant No. CTS-0210551-NER, and the Department of Energy, under Grant No. DE-FG02-02ER4600.

**Figure captions**

**Fig. 1** Illustration of a model cone. Geometry and interactions are as described in the Materials and Methods section.

**Fig. 2 a.** Simulated "magic number" clusters of cone-shaped particles from MC simulations for $N = 4 - 17$. Note $N = 8a$ is a snub-disphenoid and $N = 8b$ is a twisted-square. **b.** Predicted structures from square-well model of attractive hard spheres and "second moment" system of spherical particles, both constrained to a convex hull, for $N = 4 - 17$.

**Fig. 3 a**. Simulated "magic number" clusters of cone-shaped particles obtained from Monte Carlo simulation at $N = 20, 27, 32$ and $42$. **b.** Clusters formed from simulation of square-well hard spheres and second moment model, both with convexity constraint, for $N = 20, 27, 32, 42, 50, 72$ and $132$. Structures at $N = 12, 32, 72$ and $132$ have icosahedral symmetry corresponding to $T = 1, 3, 7$ and $13$ virus structures, respectively. The $N = 42$ cluster, though extremely similar to the $T = 4$ virus structure, does not have icosahedral symmetry. We find that for $N = 20$ and $50$ the clusters have a short cylindrical structure with local hexagonal packing. Note for $N = 132$, spherical convexity must be maintained until well beyond the estimated freezing density to achieve an ordered structure.

**Fig. 1**

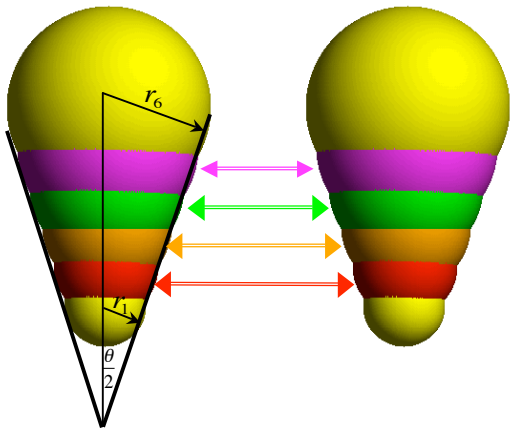

**Fig 2**

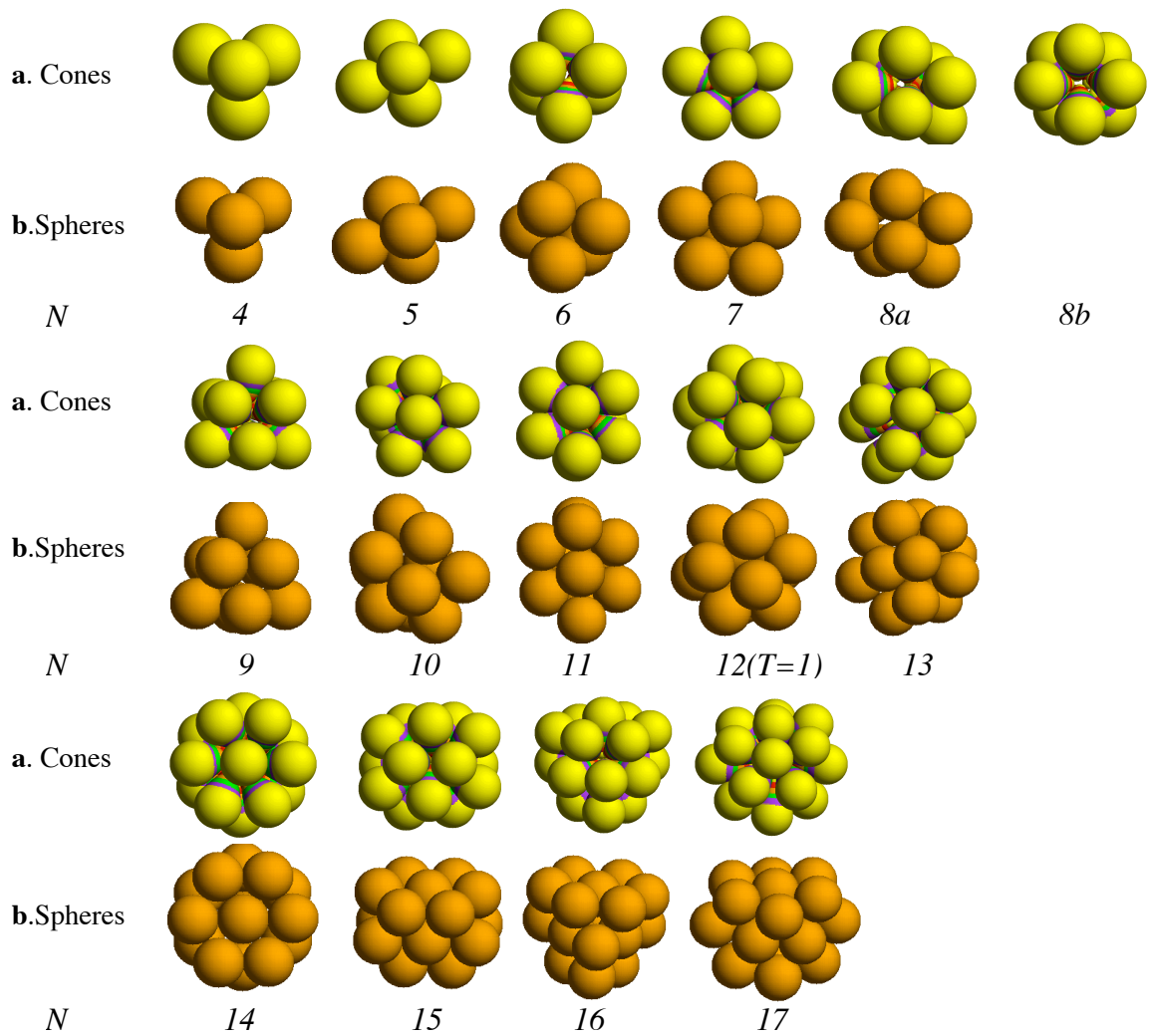

**Fig. 3**

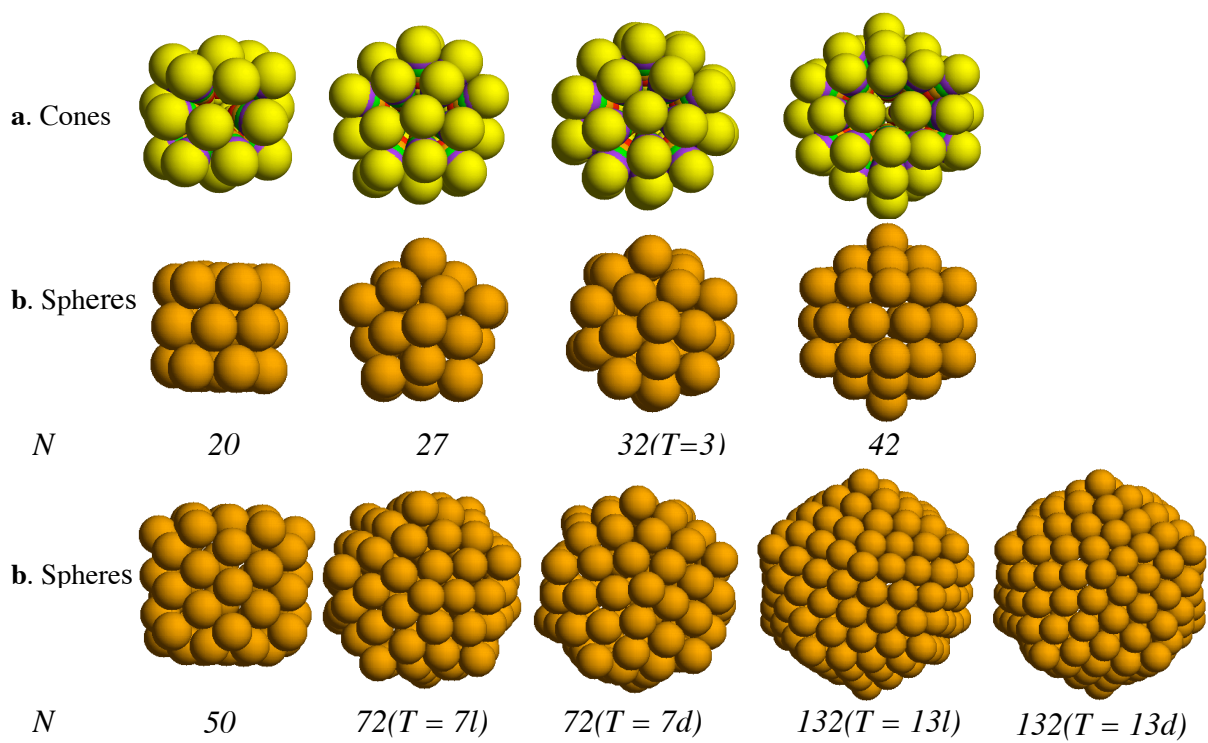

a. Cones

b. Spheres

N      20      27      32(T=3)      42

b. Spheres

N      50      72(T = 7l)      72(T = 7d)      132(T = 13l)      132(T = 13d)